\documentclass[12pt]{article}
\usepackage[utf8]{inputenc}
\usepackage[english]{babel}
\usepackage{amsfonts}
\usepackage{textcomp}
\usepackage{amsmath, amssymb, graphicx}
\renewcommand{\theequation}{\arabic{section}.\arabic{equation}}
\usepackage{geometry}
\usepackage{enumerate}
\usepackage{hyperref}
\usepackage{braket}
\usepackage{color}
\usepackage{cite}
\usepackage{float}
\usepackage{indentfirst}
\usepackage{epsfig}
\usepackage{booktabs}
\usepackage{subfigure}
\usepackage{caption}
\usepackage{ulem}
\usepackage{array}
\usepackage{multirow}

\newcommand{\be}{\begin{equation}}
\newcommand{\ee}{\end{equation}}
\newcommand{\bea}{\begin{eqnarray}}
\newcommand{\eea}{\end{eqnarray}}

\newcommand{\lb}{\label}

\newtheorem{prop}{Proposition}

\newtheorem{remark}{Remark}

\title{Continuous spin field in the $\mathbf{AdS_6}$ space}

\author{{\bf Anastasia A. Golubtsova}\footnote{\tt golubtsova@theor.jinr.ru} \; {\bf and Mikhail A. Podoinitsyn}\footnote{\tt mpod@theor.jinr.ru}}

\date{{\it Bogoliubov Laboratory of Theoretical Physics, JINR 
\\ 
141980 Dubna, Moscow Region, Russia}}

\begin{document}

\maketitle

\begin{abstract}
    A representation of the $\mathfrak{so}(2,5)$ algebra corresponding to the continuous spin field in $\mathbf{AdS_6}$ is considered. The algebra is realized using the Lie-Lorentz derivative, which naturally incorporates $\mathbf{AdS_6}$ geometry and spin degrees of freedom. Within this framework, we derive explicit expressions for the Casimir operators in terms of both the covariant derivative and the spin invariants. The continuous spin representation under consideration is defined by a system of operator constraints that generalize those known for six-dimensional Minkowski space. We demonstrate that these constraints completely fix all Casimir operators of the $\mathfrak{so}(2,5)$ algebra, with the eigenvalues determined by a dimensional real parameter $\boldsymbol{\mu}$ and a positive (half-)integer $s$.
\end{abstract}

\newpage

\tableofcontents

\newpage

\section*{Introduction}
\addcontentsline{toc}{section}{Introduction}

In higher spin theories  \cite{Snowmass,reviewsV,PL} there is a growing interest in continuous spin fields (or representations) \cite{w-1,w-2,w-3,BekSk}, particularly in spaces of constant curvature. Such representations have been extensively studied in various dimensions encompassing both flat \cite{BeBo,l-1-1,l-1-2,l-2-1,l-2-2,l-2-3,l-2-4, l-5-1,l-5-2,l-5-3,l-4-1,l-6,l-6-1,l-7-1,takata,Resh,4D-bfi-1,4D-bfi-2,4D-bfi-3,BIFP-6d,BFI-6D-tf,D6-LC,D6-LD} and curved \cite{l-3-1,l-3-2,l-3-3,l-3-4,l-3-5,l-3-6,l-3-7,l-4-2,BFIK-4D-ads-p, BFIK-4D-ads,BFIP} spacetime backgrounds\footnote{ In works \cite{l-3-6}, \cite{l-4-2}, flat and curved cases are considered together.}.

Being quite exotic for a long time, the study of continuous spin fields has naturally developed in  several complementary directions.
The Lagrangian description of free bosonic and fermionic continuous spin fields has been considered in many papers by different methods (see for e.g. \cite{l-1-2, l-2-3, l-5-2, l-4-1, takata, 4D-bfi-3, D6-LD, l-3-1,l-3-2,l-3-3, BFIK-4D-ads}). Interacting continuous spin fields have been studied in \cite{l-2-1}, \cite{l-2-4}, \cite{l-5-3}, \cite{l-6-1}, \cite{Me-int-1, Me-int-2, Me-int-3}. Group-theoretic aspects of continuous spin representations were discussed in \cite{BeBo}, \cite{l-3-5}, \cite{BFIP}, \cite{l-8,JO,WO-bipf} (see also \cite{KM}, where continuous spin representations are referred to in the framework of the dual pair correspondence). Moreover, considerable attention has been paid to supersymmetric extensions \cite{l-1-1}, \cite{SCS-N,SCS-NN, SCS-0,SCS-1,SCS-2} of continuous spin fields and continuous spin (super)particle models \cite{4D-bfi-1, 4D-bfi-2},\cite{BFIK-4D-ads-p}, \cite{JO}, \cite{SCS-1,SCS-2} in various geometric settings. As mentioned previously, the description of continuous spin fields admits multiple  approaches: frame-like \cite{l-4-1}, \cite{l-4-2}; BRST \cite{l-1-2}, \cite{takata, Resh}, \cite{BFIK-4D-ads}; light-cone \cite{l-3-4, l-3-5,l-3-6,l-3-7}; and unfolded formalism \cite{l-4-3, l-7-2}. In higher dimensions, mixed-symmetry continuous spin fields emerge, which deserve a separate investigation \cite{BeBo}, \cite{l-7-1}, \cite{l-3-3}, \cite{l-3-6}, \cite{l-4-2}.

This paper is devoted to the continuous (infinite) spin representation of the symmetry algebra $\mathfrak{so}(2,5)$ of the six-dimensional anti-de Sitter space $\mathbf{AdS_6}$. We work in a twistor formulation \cite{BFI-6D-tf}. Our approach aligns with the methodological framework established in\cite{BFIK-4D-ads-p,BFIK-4D-ads,BFIP}. The representation under consideration is realized on functions defined on the space parameterized by points on $\mathbf{AdS_6}$ and additional spinor coordinates. First, we generalize the operator constraints, which describe a continuous spin field in the six-dimensional Minkowski space \cite{D6-LC} to the $\mathbf{AdS_6}$ case. Next, we verify that the new operator constraints realize the $\mathfrak{so}(2,5)$ algebra representation and introduce an appropriate notation for the $\mathfrak{so}(2,5)$ Casimir operators, demonstrating that these constraints uniquely determine eigenvalues of all Casimirs. The resulting representation is characterized by one real parameter $\boldsymbol{\mu}$ and a positive (half-)integer parameter $s$. For $s=0$ we also compare the eigenvalues with Metsaev's classification: $\mathbf{ii}$, $\mathbf{iii}$ (when $\boldsymbol{\mu}>0$)
cases in the notation of the work \cite{l-3-5}.

Our results for $\mathbf{AdS_6}$  manifest close parallels with the $\mathbf{AdS_4}$  case, which is discussed \cite{BFIK-4D-ads,BFIP}: the operator constraints derived from algebra closure conditions fix the eigenvalues of Casimir operators, while revealing their non-independence (see \cite{most-1,most-2,most-3,most-4} and our previous discussion in \cite{BFIP} concerning the most degenerate continuous unitary representation of the $\mathbf{AdS}$  isometry group). The current work differs from the papers  on the $\mathbf{AdS_4}$ case \cite{BFIK-4D-ads,BFIP} in that it omits the discussion of unitarity,
since the crucial evidence of it, a real Lagrangian \cite{BFIK-4D-ads} constructed from similar constraints for the $\mathbf{AdS_4}$ case,
has not yet been developed for $\mathbf{AdS_6}$.

The paper is organized as follows\footnote{The first two sections consider a more general case of the $\mathbf{S_{p+1,q}}$ space, which for $p=1$ and $q=5$ is the $\mathbf{AdS_6}$ space.}.
In Section {\bf 1} we briefly review essential properties of the Casimir operators of the algebras $\mathfrak{so}(p,q)$, $\mathfrak{iso}(p,q)$, and $\mathfrak{so}(p+1,q)$. In Section {\bf 2} we discuss the geometry of pseudospheres $\mathbf{S_{p+1,q}}$, we show that the $\mathfrak{so}(p+1,q)$ generators can be expressed through Lie-Lorentz derivatives, and prove that Casimir operators can be rewritten using covariant derivatives. In Section {\bf 3} we derive operator constraints for the continuous spin representation of $\mathfrak{so}(2,5)$ and demonstrate their role in fixing all Casimir operators.

\section{Conventions for the Casimir operators}

In this section  we  discuss Casimir operators for algebras $\mathfrak{so}(p,q) \,,  \mathfrak{iso}(p,q)$ and $\mathfrak{so}(p+1,q)$. We fix the generators of these algebras, general expressions for Casimir operators and find explicit formulae for lower Casimir operators. We also demonstrate some properties of the Casimir operators of the algebras $\mathfrak{so}(p,q)$ and $\mathfrak{iso}(p,q)$ for special realizations of the corresponding generators. We discuss the case of $\mathfrak{so}(p+1,q)$ separately, since we consider this algebra as an isometry algebra of the constant curvature space $\mathbf{S_{p+1,q}}$, which is a pseudosphere embedded in $\mathbb{R}^{p+1,q}$. Therefore we choose a different kind of generators for it than for $\mathfrak{so}(p,q)$. We conclude this section by deriving an explicit formula relating the Casimir operators of these three algebras.

\subsection{Casimir operators of the $\mathfrak{so}(p,q)$ algebra}
\label{sec:so-casimir}

Our starting point is the Lie algebra $\mathfrak{so}(p,q)$, where $p + q = D$, with  generators $J_{m n}$ satisfying the commutation relations:
\begin{equation}
\label{eq:so-comm}
[J_{m n}, J_{k l}] = 
\eta_{n k} J_{m l} + \eta_{m l} J_{n k} 
- \eta_{n l} J_{m k} - \eta_{m k} J_{n l},
\end{equation}
where $m, n, k, l = 0, 1, \dots D-1$ and
 $\eta_{m n}$ is the metric of the pseudo-Euclidean space $\mathbb{R}^{p,q}$, explicitly given by:
\be\lb{flat-met-D}
||\eta_{m n}|| = \mathrm{diag}(\underbrace{-1, \dots, -1}_{p}, \underbrace{+1, \dots, +1}_{q}).
\ee

\paragraph{Explicit Realization of Generators} -- The generators $J_{m n}$ of $\mathfrak{so}(p,q)$  with \eqref{eq:so-comm} can be expressed in terms of position and momentum operators, $X^{(i)}_{m}$ and $P^{(i)}_{n}$, correspondingly. For example, in a representation with $\mathfrak{m}$ degrees of freedom, the generators $J_{mn}$ are:
\begin{equation}
\label{eq:J-explicit}
J^{(\mathfrak{m})}_{m n} = \sum_{i=1}^{\mathfrak{m}} \left(X^{(i)}_{m} P^{(i)}_{n} - X^{(i)}_{n} P^{(i)}_{m}\right).
\end{equation}
To respect  \eqref{eq:so-comm} for \eqref{eq:J-explicit},  the canonical commutation relations must hold:
\[
[P^{(i)}_{m}, X^{(j)}_{n}] = \eta_{m n} \delta^{ij}, \quad \delta^{ij} = \text{Kronecker delta},
\]
where $i,j=1,\ldots, \mathfrak{m}$.
Note that here, $\mathfrak{m} \in \mathbb{N}$ can be arbitrary, but for our purposes, it suffices to consider $1 \leq \mathfrak{m} \leq \lfloor D/2 \rfloor$, where $\lfloor x \rfloor$ denotes the integer part of $x$.

\paragraph{Casimir Operators} -- We are interested in the Casimir operators of $\mathfrak{so}(p,q)$, which are given as the following contraction of the generators:

\begin{equation}
\label{eq:casimir-def}
\mathfrak{C}_{2n} := \mathrm{c}_{2n} \, \mathcal{A}^{k_1 k_2 \dots k_{2n}}_{m_1 m_2 \dots m_{2n}} 
J_{k_1 k_2} \cdots J_{k_{2n-1} k_{2n}} J^{m_1 m_2} \cdots J^{m_{2n-1} m_{2n}},
\end{equation}
with $n = 1, \dots, \lfloor D/2 \rfloor$. In \eqref{eq:casimir-def} $\mathcal{A}$ denotes the complete antisymmetrizer taken in the form 
\be \lb{full-asym}
\mathcal{A}^{k_1 k_2 \dots k_{2n}}_{m_1 m_2 \dots m_{2n}} = \sum_{\sigma \in S_{2n}}  (-1)^{P(\sigma)} \, \delta_{m_{\sigma(1)}}^{k_1} \delta^{k_2}_{m_{\sigma(2)}} \cdots \delta^{k_{2n}}_{m_{\sigma(2n)}}\,, 
\ee
where $S_{2n}$ is a permutation group of $2n$-order and $P(\sigma)$ is the parity of the permutation $\sigma$. For convenience, the normalization constants $\mathrm{c}_{2n}$ in \eqref{eq:casimir-def} are chosen as:
\begin{equation}
\label{eq:cn-def}
\mathrm{c}_{2n} = (-1)^p \frac{1}{(2^n n!)^2}.
\end{equation}
\begin{remark} \lb{rep-for-cadab}
 Notably that the Casimir operators $\mathfrak{C}_{2\ell}$ are equal to zero, for realization  \eqref{eq:J-explicit} with $\mathfrak{m} < \ell$.
\end{remark}

\paragraph{Explicit Low-Order Casimirs} -- 
For certain small $n$ we are able to write down a compact explicit form of the Casimir operators, that we will use later. Namely,
the first three Casimir operators are reduced to:
\begin{align}
\label{eq:C2}
\mathfrak{C}_2 &= (-1)^p \left(-\frac{1}{2}J_{(2)} \right) , \\
\label{eq:C4}
\mathfrak{C}_4 &= (-1)^p \left( \frac{(D-2)(D-3)}{8} J_{(2)} + \frac{1}{8} J_{(2)}^2 - \frac{1}{4} J_{(4)} \right ), \\
\label{eq:C6}
\mathfrak{C}_6 &= (-1)^p \left( \alpha_2 J_{(2)} + \alpha_{2,2} J_{(2)}^2 + \alpha_4 J_{(4)} 
- \frac{1}{48} J_{(2)}^3 + \frac{1}{8} J_{(2)} J_{(4)} - \frac{1}{6} J_{(6)} \right ),
\end{align}
where the invariants $J_{(2k)}$ are defined as:
\begin{equation}
\label{eq:Jk-def}
J_{(2)} = J_{m n} J^{n m}, \quad 
J_{(4)} = J_{m n} J^{n k} J_{k l} J^{l m}, \quad 
J_{(6)} = J_{m n} J^{n k} \cdots J^{l  m},
\end{equation}
and the coefficients $\alpha_i$ are given by:
\begin{align}
\label{eq:alpha2}
\alpha_2 &= -\frac{11}{3} + \frac{31}{6} D - \frac{17}{6} D^2 + \frac{3}{4} D^3 - \frac{1}{12} D^4, \\
\label{eq:alpha22}
\alpha_{2,2} &= -\frac{13}{24} + \frac{13}{48} D - \frac{1}{16} D^2, \quad 
\alpha_4 = \frac{11}{6} - \frac{4}{3} D + \frac{1}{3} D^2,
\end{align}
with $D=p+q$.

\subsection{Casimir operators of the $\mathfrak{iso}(p,q)$ algebra}
\label{sec:iso-casimir}

Now we come to the inhomogeneous algebra $\mathfrak{iso}(p,q)$, which extends $\mathfrak{so}(p,q)$  by translations $P_m$, satisfying:
\begin{equation}
\label{eq:iso-comm}
[P_{m}, P_{n}] = 0, \quad 
[J_{m n}, P_{k}] = \eta_{n k} P_{m} - \eta_{m k} P_{n},
\end{equation}
along with the $\mathfrak{so}(p,q)$ algebra relations \eqref{eq:so-comm}. The indices $k,m,n$ run from $0$ to $D-1$,  as in the $\mathfrak{so}(p,q)$ case, with $D=p+q$.

\paragraph{Generalized Pauli-Lubanski Tensors} --
We can define the $(D-2n-1)$-rank  tensors:
\begin{equation}
\label{eq:W-def}
W^{(n)}_{m_1 \dots m_{D-2n-1}} = 
\varepsilon_{m_1 \dots m_{D-2n} m_{D-2n+1} \dots m_{D-1} m_D} 
J^{m_{D-2n} m_{D-2n+1}} \cdots J^{m_{D-2} m_{D-1}} P^{m_D},
\end{equation}
where $n = 0, 1, \dots, \lfloor D/2 \rfloor - 1$ and $\varepsilon_{\dots}$ is a Levi-Civita tensor. For $n=0$, we have $W^{(0)}_I = P_I$, i.e. it corresponds to translations, while for $n=1$  one gets the $D$-dimensional tensor analog of the Pauli-Lubanski pseudovector.

\paragraph{Casimir Operators} -- The Casimir operators of $\mathfrak{iso}(p,q)$ are constructed using \eqref{eq:W-def} as (see also \cite{l-1-1}):
\begin{equation}
\label{eq:C2n-def}
C_{2(n+1)} = \mathrm{n}_{2(n+1)} \, W^{(n)}_{m_1 \dots m_{D-2n-1}} W^{(n) \,  m_1 \dots m_{D-2n-1}},
\end{equation}
or, equivalently:
\begin{align}
\label{eq:C2n-antisym}
C_{2(n+1)} &= \hat{\mathrm{c}}_{2(n+1)} \, \mathcal{A}^{k_{D-2n} \dots k_D}_{m_{D-2n} \dots m_D} 
J_{k_{D-2n} k_{D-2n+1}} \cdots J_{k_{D-2} k_{D-1}} P_{k_D} \nonumber \\
&\quad \cdot J^{m_{D-2n} m_{D-2n+1}} \cdots J^{m_{D-2} m_{D-1}} P^{m_D}.
\end{align}
In \eqref{eq:C2n-def}-\eqref{eq:C2n-antisym} $\mathrm{n}_{2(n+1)}$  and $\hat{\mathrm{c}}_{2(n+1)}$
are normalization constants, given by:
\begin{equation}
\label{eq:iso-norm}
\mathrm{n}_{2(n+1)} = \frac{1}{(D-2n-1)! (2^n n!)^2}, \quad 
\hat{\mathrm{c}}_{2(n+1)} = (-1)^p \frac{1}{(2^{n} n!)^2},
\end{equation}
where the factor $(-1)^p$ arises from the convention for the Levi-Civita tensor:
\begin{equation}
\label{eq:eps-convention}
\varepsilon_{0 \dots D-1} = 1, \quad 
\varepsilon^{0 \dots D-1} = (-1)^p.
\end{equation}

\paragraph{Explicit Low-Order Casimirs} -- As for the $\mathfrak{so}(p,q)$ algebra we can write down the compact explicit formulae for the Casimir operators for certain small $n$. Thus,
the first three Casimir operators are represented by (see \cite{BIFP-6d}):
\begin{align}
\label{eq:C2-iso}
C_2 &= (-1)^p P_m P^m, \\
\label{eq:C4-iso}
C_4 &= (-1)^{p+1} \left(\Pi_m \Pi^m + \frac12 J_{(2)} P_m P^m\right), \\
\label{eq:C6-iso}
C_6 &= (-1)^p \Bigl( \Pi^{m} J_{m n} \Pi_{k} J^{k n} 
+ \frac{1}{8} J_{(2)}^2 P_m P^m - \frac{1}{4} J_{(4)} P_m P^m \nonumber \\
&\quad + \frac{1}{2} J_{(2)} \Pi_m \Pi^m + \beta_1 J_{(2)} P_m P^m + \beta_2 \Pi_m \Pi^m \Bigr),
\end{align}
where $\Pi_n$ is defined as:
\begin{equation}
\label{eq:Pi-def}
\Pi_{n} = P^{m} J_{m n},
\end{equation}
and the constants $\beta_i$ in \eqref{eq:C6-iso} are:
\begin{equation}
\label{eq:beta-def}
\beta_1 = \frac{1}{8} \left(D^2 - 9D + 10\right), \quad \beta_2 = -(D-2).
\end{equation}

\paragraph{Casimir operators in terms of spin generators} --
In the end of this subsection, we demonstrate that decomposing the $\mathfrak{so}(p,q)$ subalgebra generators into orbital and spin parts, the orbital component cancels out in the Casimir operators. Although this follows directly from the structure of the Casimir operators \eqref{eq:C2n-antisym}, we also provide explicit expressions illustrating this fact for our further use. A similar property holds for the isometry algebra of the $\mathbf{S_{p+1,q}}$ space.

First, let us decompose the $\mathfrak{so}(p,q)$ generators into the orbital and spin parts\footnote{The formula \eqref{eq:J-decomp} with \eqref{eq:M-comm} can be considered as some realization of  $\mathfrak{iso}(p,q)$, where $X_n$ are coordinates on $\mathbb{R}^{p,q}$, $P_n$ is a momentum, i.e. $P_n = \partial/\partial X^n $, and the spin generators $\mathcal{M}_{m n} $ can be realized in various ways.}:
\begin{equation}
\label{eq:J-decomp}
J_{m n} = X_{m} P_{n} - X_{n} P_{m} + \mathcal{M}_{m n},
\end{equation}
where the spin part $\mathcal{M}_{m n}$ and the operators $X_m, P_n$ satisfy the commutator relations:
\begin{align}
\label{eq:M-comm}
[P_{m}, X_{n}] &= \eta_{m n}, \quad 
[X_{m}, X_{n}] = [\mathcal{M}_{m n}, X_{k}] = [\mathcal{M}_{m n}, P_{k}] = 0, \nonumber \\
[\mathcal{M}_{m n}, \mathcal{M}_{k l}] &= 
\eta_{n k} \mathcal{M}_{m l} + \eta_{m l} \mathcal{M}_{n k} 
- \eta_{n l} \mathcal{M}_{m k} - \eta_{m k} \mathcal{M}_{n l}.
\end{align}

Then substituting \eqref{eq:J-decomp} into \eqref{eq:C4-iso}--\eqref{eq:C6-iso}, we find the relations for the Casimirs $C_4$ and $C_6$, for which the orbital part vanish:
\begin{align}
\label{eq:C4-spin}
C_4 &= (-1)^{p+1} \left(\mathcal{P}_m \mathcal{P}^m + \frac12 \mathcal{M}_{(2)} P_m P^m\right), \\
\label{eq:C6-spin}
C_6 &= (-1)^p \Bigl( \mathcal{P}^{m} \mathcal{M}_{m n} \mathcal{P}_{k} \mathcal{M}^{k n} 
+ \frac{1}{8} \mathcal{M}_{(2)}^2 P_m P^m - \frac{1}{4} \mathcal{M}_{(4)} P_m P^m \nonumber \\
&\quad + \frac{1}{2} \mathcal{M}_{(2)} \mathcal{P}_m \mathcal{P}^m
+ \boldsymbol{\beta}_1 \mathcal{M}_{(2)} P_m P^m + \boldsymbol{\beta}_2 \mathcal{P}_m \mathcal{P}^m \Bigr).
\end{align}
In \eqref{eq:C4-spin}-\eqref{eq:C6-spin}
$\mathcal{P}_m$ and $\mathcal{M}_{(2n)}$ are defined analogously to \eqref{eq:Pi-def} and \eqref{eq:Jk-def} (with the replacement: $J_{m n} \to \mathcal{M}_{m n}$) and the coefficients $\boldsymbol{\beta}_i$ in \eqref{eq:C6-spin} are:
\begin{equation}
\label{eq:beta-spin}
\boldsymbol{\beta}_1 = \frac{1}{8} \left(D^2 - 5D + 10\right), \quad \boldsymbol{\beta}_2 = 1.
\end{equation}

\subsection{The $\mathfrak{so}(p+1,q)$ algebra and its flat limit $\mathfrak{iso}(p,q)$}
\label{sec:so-iso-limit}

In this section we consider the algebra $\mathfrak{so}(p+1,q)$ as the isometry algebra of a pseudosphere $\mathbf{S_{p+1,q}}$ in $\mathbb{R}^{p+1,q}$ characterized by the radius $R$. By making some standard redefinition of the generators of $\mathfrak{so}(p+1,q)$ using the parameter $R$ we obtain  the Casimir operators in terms of new generators. This is done in order to explicitly demonstrate   the well-known group contraction \cite{IW}, that we will use below. This procedure at the level of algebras implies that as $R \to \infty$ the algebra $\mathfrak{so}(p+1,q)$ turns into $\mathfrak{iso}(p,q)$.

The embedding  of  the hypersurface  $\mathbf{S_{p+1,q}}$ into the  $\mathbb{R}^{p+1,q}$ space is  defined by (see for e.g. \cite{IR}):
\begin{equation}
\label{eq:sphere-def}
X^A X^B \eta_{AB} = -R^2, \quad A,B = 0, \dots, D,
\end{equation}
with the metric:
\begin{equation}
\label{eq:eta-AB}
||\eta_{AB}|| = \mathrm{diag}(\underbrace{-1, \dots, -1}_{p}, \underbrace{+1, \dots, +1}_{q}, -1)
\end{equation}
and $D=p+q$.
The connected component of the identity of the symmetry group of $\mathbf{S_{p+1,q}}$ is $SO(p+1,q)$, whose Lie algebra generators\footnote{For $J_{A B}$ we use commutation relations as \eqref{eq:so-comm} with the replacement: $J_{m n} \to J_{A B}$ and $\eta_{m n} \to \eta_{A B}$. } $J_{AB}$ can be decomposed as:
\begin{equation}
\label{eq:so-split}
P_n = R^{-1} J_{nD}, \quad J_{mn}, \quad m,n = 0, \dots, D-1.
\end{equation}
Note that the generators $P_n$ and $J_{mn}$ obey:
\begin{align}
\label{eq:so-comm-split}
[P_n, P_m] &= R^{-2} J_{nm}, \quad 
[J_{mn}, P_l] = \eta_{nl} P_m - \eta_{ml} P_n, \nonumber \\
[J_{mn}, J_{lk}] &= \eta_{nl} J_{mk} + \eta_{mk} J_{nl} - \eta_{nk} J_{ml} - \eta_{ml} J_{nk} \,, 
\end{align}
where $\eta_{m n}$ was defined in \eqref{flat-met-D}. 
\paragraph{Flat Limit} --
In the limit $R \to \infty$, the hypersurface $\mathbf{S_{p+1,q}}$ turns into  the pseudo-Euclidean space $\mathbb{R}^{p,q}$, so the $\mathfrak{so}(p+1,q)$  algebra contracts \cite{IW} to $\mathfrak{iso}(p,q)$.

The Casimir operators of $\mathfrak{so}(p+1,q)$ are related to those of $\mathfrak{so}(p,q)$ and $\mathfrak{iso}(p,q)$ through the following relation (see a similar expression in \cite{LJ}):
\begin{equation}
\label{eq:casimir-limit}
\mathcal{C}_{2n} = R^2 \, \widehat{C}_{2n} - \mathfrak{C}_{2n},
\end{equation}
where the hat in $\widehat{C}_{2n}$ denotes that  $\widehat{C}_{2n}$ has the same form as R.H.S of \eqref{eq:C2n-def}, but with non-selfcommuting operators $P_n$. Thus, in the flat limit we have 
\be \lb{flat-l-cas}
[P_n, P_m] \overset{R \to \infty}{=} 0 \quad \Longrightarrow \quad \mathcal{C}_{2n}/R^2 \overset{R \to \infty}{=} C_{2n} \,. 
\ee

To derive the relation \eqref{eq:casimir-limit}, we first apply the general formula for the Casimir operator  \eqref{eq:casimir-def} to the  generators $J_{AB}$ of $\mathfrak{so}(p+1,q)$. Implementing the generator decomposition from \eqref{eq:so-split}, and doing some algebra we are brought to the relation \eqref{eq:casimir-limit}.

\paragraph{Explicit Low-Order Casimirs} --
The first three Casimir operators of the $\mathfrak{so}(p+1,q)$ algebra are:
\begin{align}
\label{eq:C2-AdS}
\mathcal{C}_2 &= (-1)^p \left(R^2\, P_n P^n + \frac12 J_{(2)} \right) , \\
\label{eq:C4-AdS}
\mathcal{C}_4 &= (-1)^{p+1}\, R^2 \left(\Pi_n \Pi^n + \frac12 J_{(2)} P_n P^n\right) + O_4(J), \\
\label{eq:C6-AdS}
\mathcal{C}_6 &= (-1)^p \,R^2\, \Bigl( \Pi^{n} J_{n m} \Pi_{k} J^{k m} 
+ \frac{1}{8} J_{(2)}^2 P_n P^n - \frac{1}{4} J_{(4)} P_n P^n \nonumber \\
&\quad + \frac{1}{2} J_{(2)} \Pi_n \Pi^n + \beta_1 J_{(2)} P_n P^n + \beta_2 \Pi_n \Pi^n \Bigr) - O_6(J),
\end{align}
where we define
\begin{align} \label{rel-O4}
O_4(J) = (-1)^{p+1} \left (- \frac{(-2 + 5 D - D^2)}{8} J_{(2)} + \frac18 J_{(2)}^2  -\frac14 J_{(4)} \right)\\
\label{rel-O6}
O_6(J) =(-1)^{p} \left(\gamma_2 J_{(2)} + \gamma_{2,2} J_{(2)}^2 + \gamma_4 J_{(4)} - \frac{1}{48} J_{(2)}^3 + \frac{1}{8} J_{(2)} J_{(4)} - \frac16 J_{(6)}\right) 
\end{align} 
and the constants $\gamma_i$ in \eqref{rel-O6} are fixed as
\begin{align}
\label{eq:gamma2}
\gamma_2 &= - \frac{1}{12} \left(D^4-9 D^3+28 D^2-32 D+8\right), \\
\label{eq:gamma22}
\gamma_{2,2} &= \frac{1}{48} \left(-3 D^2+13 D-14\right), \quad 
\gamma_4 = \frac{1}{6} \left(2 D^2-8 D+5\right).
\end{align}
The constants $\beta_i$  in \eqref{eq:C6-AdS} match with  those \eqref{eq:beta-def} as it should occur
  for the correct flat limit \eqref{flat-l-cas}.

Our next step will be devoted to simplifying  the Casimirs given by \eqref{eq:C2-AdS}-\eqref{eq:C6-AdS} for some realization of the $\mathfrak{so}(p+1,q)$ algebra. 
In other words, we want to obtain some analogue of the  transformation from \eqref{eq:C4-iso}-\eqref{eq:C6-iso} to \eqref{eq:C4-spin}-\eqref{eq:C6-spin} for the isometry algebra of the $\mathbf{S_{p+1,q}}$ space.

\setcounter{equation}{0}

\section{$\mathbf{S_{p+1,q}}$ space: geometry and algebra}
In this section  we express  the Casimir operators \eqref{eq:C2-AdS}-\eqref{eq:C6-AdS} of the isometry algebra of $\mathbf{S_{p+1,q}}$ using  geometric  quantities, that significantly simplifies further analysis devoted to the representation of continuous spin in $\mathbf{AdS_6}$ (in our notation it corresponds to the $\mathbf{S_{1+1,5}}$ space).

First, we fix the notation for  geometric quantities that we will use. Then, we choose certain metric for $\mathbf{S_{p+1,q}}$ and calculate geometric objects  of the space. Next, we write a realization for the generators $\mathfrak{so}(p+1,q)$ in terms of the geometric objects.

We are looking for a realization of the generators $P_n$ and $J_{m n}$ defined by \eqref{eq:so-split}, on the space of functions $\Psi(x^{\mu}, \mathfrak{y})$, where $x^{\mu}$ are coordinates on $\mathbf{S_{p+1,q}}$, and $\mathfrak{y}$ is a set of additional variables (it can be  either $\mathfrak{so}(p,q)$ vector or spinor) related to the spin degrees of freedom\footnote{By introducing such additional variables we come from tensor fields to generating functions (see for e.g. \cite{PL}), which  simplifies drastically the analysis.}, which we do not  specify up to the last section.
If we consider the functions $\Psi(x^{\mu})$ instead of $\Psi(x^{\mu}, \mathfrak{y})$, then the realization of the generators is given by the Lie derivatives along the corresponding Killing vectors.
 In our case, the generators will be expressed through the so-called Lie-Lorentz derivative, which was originally constructed  to realize the symmetry algebra on spinors in the curved space in \cite{YK} and \cite{WGU}. It is interesting to note that at the moment the concept of the Lie-Lorentz derivative often arises in various supersymmetric theories (see \cite{nLLD, VL} and references therein, see also \cite{PanKe} where this derivative is used in higher spin theories).

In addition to the direct derivation of the generators $\mathfrak{so}(p+1,q)$ in terms of the Lie-Lorentz derivative, we briefly present three more seemingly different methods for obtaining the corresponding generators: {\bf a)} the first is based on a simple observation about the calculation of the commutator of operators close in structure to the covariant derivative; { \bf b)} the second method is associated with the so-called ambient formalism (see \cite{PL} and references therein); {\bf c)} the third method is based on the fact that $\mathfrak{so}(p+1,q)$ is a subalgebra of the conformal algebra
 $\mathfrak{conf}(\mathbb{R}^{p,q})$.

Next, for the obtained representation, we write down a number of $\mathfrak{so}(p+1,q)$ invariant operators through which we express the Casimir operators. 

\subsection{Geometrical objects of the $\mathbf{S_{p+1,q}}$ space} \lb{ADS4C}

The metric tensor $g_{\mu \nu}(x)$ for $\mathbf{S_{p+1,q}}$ defined by \eqref{eq:sphere-def} is chosen in the following form (see for e.g. book \cite{IR})
\be \lb{AdSm}
g_{\mu \nu}(x) = G^{-2}(x) \eta_{\mu \nu} \,, \quad G(x)=\left(1-\frac{(x)^2}{4 R^2}\right) \,,
\ee
where $\mu, \nu = 0,1, \dots D-1$ are  wold indices, $(x)^2 := x^{\mu} x^{\nu} \eta_{\mu \nu}$ and 
\begin{equation}
\label{eq:etaSP-AB}
||\eta_{\mu \nu}|| = \mathrm{diag}(\underbrace{-1, \dots, -1}_{p}, \underbrace{+1, \dots, +1}_{q}) \,.
\end{equation}
Then the corresponding frame fields\footnote{See all notation of the geometrical objects in App. \textbf{\ref{appGeom}
}.}  are taken as 
\be \lb{FrFia}
e^{\mu}_{n} = G(x) \delta^{\mu}_{n} \,, \quad e_{\mu}^{n} = G^{-1}(x) \delta_{\mu}^{n} \,.
\ee
Thus, the vielbein $e_n$ defined in \eqref{frf} has the following explicit form
\be \lb{frfAds}
e_n = \left(1 - \frac{(x)^2}{4 R^2}\right) \partial_n \,, \qquad \partial_n := \delta^{\mu}_{n} \frac{\partial}{\partial x^{\mu}} \,. 
\ee
Taking into account the metric \eqref{AdSm} and frame fields \eqref{FrFia} we can find the anholonomy coefficient, the spin connection, the Riemann and torsion tensors, correspondingly,
\bea \lb{AC-SC-ads}
\mathcal{E}_{n m}^{\;\;\;\;\;l}  &=& - \frac{1}{2 R^2} \left ( x_{n} \delta_{m}^{l} - x_{m} \delta_{n}^{l} \right ) \,, \quad w_{k n m} = - \frac{1}{2 R^2} \left (x_n \eta_{m k} - x_{m} \eta_{n k} \right ) \,, \\[7pt] 
\lb{tR-tT-ads}
\mathcal{R}_{n m}^{\;\;\;\;\;k l} &=& -\frac{1}{R^2} \left(\delta_{n}^{k} \delta_{m}^{l}-\delta_{m}^{k} \delta_{n}^{l} \right ) \,, \quad \mathcal{T}_{n m}^{\;\;\;\;\; l} = 0 \,,
\eea
where we used \eqref{frf}-\eqref{sc-ac} and 
\be \lb{defxnxn}
x_n := \eta_{n \mu} x^{\mu} \,, \quad x^n := \delta^n_{\mu} x^{\mu} \quad \Longleftrightarrow \quad \partial_n x_m = \eta_{n m}.
\ee  Alternatively, we can find the spin connection using the condition  of covariant constancy of the frame fields
\be \lb{cd-ff}
\mathcal{D}_{\mu} e^{\nu}_n \equiv \partial_{\mu} e^{\nu}_n + \Gamma^{\nu}_{\mu \lambda} e^{\lambda}_{n} + w_{\mu n}^{\;\;\;\; m} e^{\nu}_{m} = 0,
\ee
 where $\mathcal{D}_{\mu}$ is a covariant derivative. Resolving this equation with respect to $w_{knm}$, one gets (see for e.g. \cite{PN})
\be \lb{sc-wt}
w_{k n m} = e^{\mu}_{k} \, e_{n}^{\rho} \, g_{\rho \nu} \, (\partial_{\mu} e^{\nu}_{m} + \Gamma^{\nu }_{\mu \lambda} \, e^{\lambda}_{m}) \,,
\ee
where $\Gamma^{\nu }_{\mu \lambda}$ are  Christoffel symbols  given by
\be \lb{Ch-symb}
\Gamma^{\nu}_{\mu \lambda} = \frac{1}{2 R^2 G} \left ( G^2 \, x_{\mu} \, \delta^{\nu}_{\lambda} +   G^2 \, x_{\lambda } \, \delta^{\nu}_{\mu}  - x^{\nu} \eta_{\mu \lambda} \right ) \,. 
\ee
 
\subsection{Generators of the symmetry algebra} \lb{susec-gen}

 Consider the operators $J_{m n}$ with the commutation relations \eqref{eq:so-comm} (below we will give its explicit form) and require the following commutators are fulfilled
\bea \lb{s-f-3}
[J_{nm}, e_{l}] &=& \eta_{ml} e_n - \eta_{n l} e_m \,, \\[7pt]
\lb{s-f-4}
[J_{nm}, w_{lkr}] &=& \eta_{ml}w_{nkr} - \eta_{nl}w_{mkr}+\eta_{mk}w_{lnr}-\eta_{nk}w_{lmr} +\eta_{mr}w_{lkn}-\eta_{nr}w_{lkm} \,,
\eea
where $e_{n}$ is given by \eqref{frf}, $w_{l k r}$  is the spin connection and the metric is not fixed a priori.
In addition, we consider the  operators $P_n$, which have the form
\be \lb{s-f-1}
P_n = e_n - \frac12 w_{n k l} J^{k l} \,. 
\ee
One can check that for the $P_n$ and $J_{mn}$ operators the following commutators hold 
\be \lb{s-f-2}
[P_n, P_m] = \mathcal{T}_{n m}^{\;\;\;\;\;l} P_l - \frac12 \mathcal{R}_{n m}^{\;\;\;\;\;k l} J_{k l} \,,  \qquad [J_{m n}, P_l] = \eta_{n l} P_{m} - \eta_{m l} P_{n} \, . 
\ee
where we used \eqref{eq:so-comm}, \eqref{s-f-3}-\eqref{s-f-4},  properties of the spin-connection and the torsion $\mathcal{T}_{n m}^{\;\;\;\;\;l}$ and the Riemann tensor $\mathcal{R}_{n m}^{\;\;\;\;\;kl}$ are defined by \eqref{totT}-\eqref{RtCA}, correspondingly. 

Then plugging the Riemann and the torsion tensors \eqref{tR-tT-ads}, calculated on the metric \eqref{AdSm}, into \eqref{s-f-2} and comparing the result with \eqref{eq:so-comm-split} we  see that operators $P_n$ and $J_{m n}$ form the $\mathfrak{so}(p+1,q)$ algebra. At the same time, one can explicitly present the generators $P_n$ and $J_{m n}$,  taking into account \eqref{frfAds}-\eqref{AC-SC-ads} for \eqref{s-f-1}, and considering the generators $J_{m n}$ in the form
\be \lb{ex-gen-sopq}
J_{m n} = \mathcal{L}_{m n} + \mathcal{M}_{m n} \,,  \quad \mathcal{L}_{m n} := x_{m} \partial_{n} - x_{n} \partial_{m} \,.
\ee
In \eqref{ex-gen-sopq} the spin part $\mathcal{M}_{m n}$ of  $J_{m n}$ satisfies relations  \eqref{eq:M-comm} and we also use \eqref{defxnxn}.

\begin{remark} \lb{rem-1}
In what follows below, unless otherwise stated, we will consider everywhere the case of the $\mathbf{S_{p+1,q}}$ space with the metric \eqref{AdSm}, thus the  geometric quantities are given by \eqref{FrFia}-\eqref{tR-tT-ads}.
\end{remark}

\begin{remark} \lb{rem-2}
 The generators  $P_{n}$ and $J_{m n}$
of $\mathfrak{so}(p+1,q)$ given by  \eqref{s-f-1} and \eqref{ex-gen-sopq} (taking into account Remark \textbf{\ref{rem-1}}) act on the functions $\Psi(x^{\mu}, \mathfrak{y})$, where $x^{\mu}$ are coordinates on $\mathbf{S_{p+1,q}}$ and $\mathfrak{y}$ is a set of additional variables, which depend on a realization of the spin part $\mathcal{M}_{m n}$.
\end{remark}

\paragraph{Ambient formalism and Conformal algebra} -- The generators $P_n$ defined by \eqref{s-f-1} with \eqref{frfAds}-\eqref{AC-SC-ads} and $J_{m n}$ given by \eqref{ex-gen-sopq} can be obtain using a simple generalization of the construction,  which follows from the ambient formalism (see for e.g. \cite{PL} and refs. therein) for the $\mathbf{AdS}$ symmetry algebra (see the detailed derivation for the $\mathbf{AdS_4}$ case in \cite{BFIP}).

It is worth to be noted that the same generators can be also constructed from the Lie algebra relation (see the book \cite{IR} and also the work \cite{Band})
\be \lb{LA-R}
\mathfrak{so}(p+1,q) \subset \mathfrak{so}(p+1,q+1) \simeq \mathfrak{conf}(\mathbb{R}^{p,q}) \,. 
\ee
In this approach, within the explicit construction \eqref{LA-R}, we can express $P_n$ and $J_{m n}$ using the known realization of the generators for $\mathfrak{conf}(\mathbb{R}^{p,q})$. This realization of the conformal algebra contains  also the conformal dimension parameter, in addition to the spin generators. To obtain the $\mathfrak{so}(p+1,q)$ generators written down earlier, the conformal dimension parameter must be set to zero.

\paragraph{Lie derivative and Killing vectors} -- Let us consider the case when the spin part of the angular momentum vanishes: $\mathcal{M}_{m n}= 0$. Then the generators $P_{n}$ can be represented in the form 
\be\label{PnLie}
P_n = \phi^{\mu}_{(n)} \partial_{\mu}\equiv \mathcal{L}_{\phi_{(n)}},
\ee
where $\mathcal{L}_{\phi_{(n)}}$ is a Lie derivative
and  the vectors $\phi^{\mu}_{(n)}$ are  Killing vectors of  $\mathbf{S_{p+1,q}}$, i.e. they obey the equation
\be \lb{kill-eq}
\phi^{\gamma}_{(n)} \partial_{\gamma} g_{\mu \nu} + g_{\gamma \nu} \partial_{\mu} \phi^{\gamma}_{(n)} + g_{\gamma \mu} \partial_{\nu} \phi^{\gamma}_{(n)} = 0 \,
\ee
or, equivalently,
\be
\mathcal{D}_{\mu}\phi_{\nu\,(n)}+\mathcal{D}_{\nu}\phi_{\mu\,(n)}=0,
\ee
where $\phi_{\mu\,(n)} = g_{\mu \nu}
 \, \phi^{\nu}_{(n)}$ and $\mathcal{D}_{\mu}$ acts on $\phi^{\nu}_{(n)}$ as
\be \lb{cd-kv}
\mathcal{D}_{\mu}\phi^{\nu}_{(n)}=\partial_{\mu}\phi^{\nu}_{(n)}+\Gamma^{\nu}_{\mu\lambda}\phi^{\lambda}_{(n)},
\ee
with $\Gamma^{\nu}_{\mu\lambda}$ are given by \eqref{Ch-symb}. Note that  the lower index $(n)$ of $\phi^{\nu}_{(n)}$ just labels the Killing vectors (compare \eqref{cd-kv} with \eqref{cd-ff}). The property \eqref{PnLie} was discussed in \cite{PN}.

\paragraph{Lie-Lorentz derivative} -- Now we turn to  a generic case with non-zero $\mathcal{M}_{mn}$. The Lie-Lorentz derivative (see \cite{nLLD} and refs. therein) acting on the functions $\Psi(x^{\mu},\mathfrak{y})$  has the following form 
\be \lb{lie-lor}
\mathbb{L}_{\phi_{(n)}}= \phi^{\mu}_{(n)} \mathcal{D}_{\mu} + \frac12 g_{\lambda \nu} \, \bigl(\mathcal{D}_{\mu} \, \phi^{\nu}_{(n)}\bigr) \, e^{\mu}_{a} e^{\lambda}_{b} \mathcal{M}^{a b} \,,
\ee
where $\phi^{\mu}_{(n)}$ are the Killing vectors of $\mathbf{S_{p+1,q}}$ and were introduced in the previous paragraph. In  \eqref{lie-lor} the covariant derivative in the first term  is defined as 
\be
\mathcal{D}_{\mu}=e^{n}_{\mu} \, \mathcal{D}_n,
\ee
with $\mathcal{D}_n$  given by \eqref{s-f-5} and for the second term $\mathcal{D}_{\mu}$ acts as in \eqref{cd-kv}.

One can check that the Lie-Lorentz derivative $\mathbb{L}_{\phi_{(n)}}$  is exactly the operator $P_n$, i.e. the equality 
\be
 P_n=\mathbb{L}_{\phi_{(n)}}
\ee
is valid, where $P_n$ is given by \eqref{s-f-1} with \eqref{frfAds}-\eqref{AC-SC-ads} and  \eqref{ex-gen-sopq}. 

Although in the last two paragraphs we discussed only the generators $P_n$, all these arguments can be equally applied to the generators $J_{mn}$. It is also important to note that the operator $\mathbb{L}_{\phi_{(n)}}$ \eqref{lie-lor} is valid for any choice of $\mathbf{AdS}$ metric, while the operator $P_n$ \eqref{s-f-1} appears to hold only for stereographic coordinates of $\mathbf{AdS}$.

\subsection{Covariant derivative and $\mathfrak{so}(p+1,q)$ invariants}

In this subsection we introduce the covariant derivative on the functions $\Psi(x^{\mu}, \mathfrak{y})$, which  discussed in Remark \textbf{\ref{rem-2}}. Then we will show that the Casimir operators \eqref{eq:C2-AdS}-\eqref{eq:C6-AdS} on the space of these functions can be expressed only in terms of the covariant derivative and the spin part of the angular momentum $\mathcal{M}_{m n}$ from \eqref{ex-gen-sopq}.

\paragraph{Transformation of the covariant derivative} -- First, we consider a covariant derivative defined by
\be \lb{s-f-5}
\mathcal{D}_n = e_n + \frac12 w_{n k l} \mathcal{M}^{kl} \,,
\ee
which in general (not only for the $\mathbf{S_{p+1,q}}$ case) automatically satisfies 
\be \lb{s-f-6D}
[\mathcal{D}_n, \mathcal{D}_m] = \mathcal{E}_{n m}^{\;\;\;\;\; l} \mathcal{D}_l +
\frac12 \mathcal{R}_{n m}^{\;\;\;\;\;k l} \mathcal{M}_{k l}.
\ee
\begin{prop} Let us consider the case of the $\bf{S_{p+1,q}}$ space, for which we fix the frame fields as 
\eqref{FrFia}. Then we have the following relations\footnote{In the notation of the paper \cite{nLLD} the first relation in \eqref{s-f-11} is the conservation of the covariant derivative.} 
\be\lb{s-f-11}
[P_k, \mathcal{D}_m] = w_{k m a} \mathcal{D}^a \,, \qquad 
[J_{mn}, \mathcal{D}_l] = \eta_{nl} \mathcal{D}_{m} - \eta_{ml} \mathcal{D}_n \,. 
\ee
\textbf{Proof:} The proof is a direct calculation.
\end{prop}
\paragraph{Invariant operators} -- Based on \eqref{s-f-5} we can construct a chain of operators:
\bea
\lb{LBO}
\Box &:=& \eta^{m n} (e_m e_n + w_{m n l} e^l) = \frac{1}{\sqrt{-g}} \, \partial_{\mu} \sqrt{-g}\,  g^{\mu \nu} \partial_{\nu} \,, \\[7pt]
\lb{LBO-0}
(\mathcal{D})^2 &:=& \eta^{m n} (\mathcal{D}_m \mathcal{D}_n + w_{m n l} \mathcal{D}^l) = \frac{1}{\sqrt{-g}} \, \mathcal{D}_{\mu} \sqrt{-g}\,  g^{\mu \nu} \mathcal{D}_{\nu}\,, \\[7pt]
\lb{LBO-1}
(\mathcal{D}_{\!_{\bf{(1)}}})^2 &:=& \mathcal{M}^{k m} \mathcal{M}_k^{\;\;n} (\mathcal{D}_m \mathcal{D}_n + w_{m n l} \mathcal{D}^l)  \,, \\[7pt] 
\lb{LBO-2}
(\mathcal{D}_{\!_{\bf{(2)}}})^2 &:=& \mathcal{M}^k_{\;\; d}\mathcal{M}^{d m} \mathcal{M}_{k r} \mathcal{M}^{r n} (\mathcal{D}_m \mathcal{D}_n + w_{m n l} \mathcal{D}^l) \,, \\[7pt] 
\nonumber
& \vdots & \\[7pt] 
\lb{LBO-n}
(\mathcal{D}_{\!_{\bf{(\mathrm{n})}}})^2 &:=& \Bigl(\prod\limits_{i=1}^{\mathrm{n}}\mathcal{M}\Bigr)^{k m} \Bigl(\prod\limits_{j=1}^{\mathrm{n}}\mathcal{M}\Bigr)_k^{\;\;n}(\mathcal{D}_m \mathcal{D}_n + w_{m n l} \mathcal{D}^l) \,.
\eea
The first operator \eqref{LBO} in the  chain is the Laplace-Beltrami operator, and the rest ones are some of its spin generalizations\footnote{In work \cite{BFIP} notation $(\Pi)^2$ is used for $(\mathcal{D}_{\!_{\bf{(1)}}})^2$}. The number of the operators \eqref{LBO}-\eqref{LBO-n}, i.e. $\mathrm{n}+1$, coincides with the number of Casimir operators of the $\mathfrak{so}(p+1,q)$ algebra.

\begin{remark}
It is well known that operators $\Box$ and $(\mathcal{D})^2$ are the $\mathfrak{so}(p+1,q)$ invariant. 
\end{remark}

\begin{prop} The operators $(\mathcal{D}_{\!_{\bf{(i)}}})^2$, with $i = 1, 2, \dots ,\mathrm{n}$ are also invariant under the $\mathfrak{so}(p+1,q)$ transformations, i.e. we have
\be\lb{s-f-11i}
[P_k, (\mathcal{D}_{\!_{\bf{(i)}}})^2] =0 \,, \qquad 
[J_{mn}, (\mathcal{D}_{\!_{\bf{(i)}}})^2] = 0 \,, 
\ee
\textbf{Proof:} The second eq. is obvious, because $(\mathcal{D}_{\!_{\bf{(i)}}})^2$ is a  scalar under Lorentz transformations (see 2nd eq. from \eqref{s-f-11}). The proof of the first eq. in \eqref{s-f-11i} is a direct calculation. 
\end{prop}

\paragraph{Low order Casimirs of $\mathfrak{so}(p+1,q)$ in terms of the covariant derivative} --
Substituting \eqref{s-f-1} in \eqref{eq:C2-AdS}-\eqref{eq:C6-AdS} with \eqref{AC-SC-ads}-\eqref{tR-tT-ads} we get 
\begin{align}
\label{eq:C2-AdS-D}
\mathcal{C}_2 &= (-1)^p \left(R^2\, (\mathcal{D})^2 + \frac12 \mathcal{M}_{(2)} \right) , \\
\label{eq:C4-AdS-D}
\mathcal{C}_4 &= (-1)^{p+1}\, R^2 \left((\mathcal{D}_{_{\bf{(1)}}})^2 + \frac12 \mathcal{M}_{(2)} (\mathcal{D})^2 \right) + \mathcal{O}_4(\mathcal{M}), \\
\label{eq:C6-AdS-D}
\mathcal{C}_6 &= (-1)^p \,R^2\, \Bigl( (\mathcal{D}_{_{\bf{(2)}}})^2 
+ \frac{1}{8} \mathcal{M}_{(2)}^2 (\mathcal{D})^2 - \frac{1}{4} \mathcal{M}_{(4)} (\mathcal{D})^2 \nonumber \\
&\quad + \frac{1}{2} \mathcal{M}_{(2)} (\mathcal{D}_{_{\bf{(1)}}})^2 + \boldsymbol{\beta}_1 \mathcal{M}_{(2)} (\mathcal{D})^2 + \boldsymbol{\beta}_2 (\mathcal{D}_{_{\bf{(1)}}})^2 \Bigr) - \mathcal{O}_6(\mathcal{M}),
\end{align}
where we introduce
\begin{align} \label{rel-O4-D}
\mathcal{O}_4(\mathcal{M}) &= (-1)^{p+1} \left (\frac{(D-2)(D+1)}{8} \mathcal{M}_{(2)} + \frac18 \mathcal{M}_{(2)}^2  -\frac14 \mathcal{M}_{(4)} \right)\\
\label{rel-O6-M}
O_6(\mathcal{M}) &=(-1)^{p} \Bigl(\boldsymbol{\gamma}_2 \mathcal{M}_{(2)} + \boldsymbol{\gamma}_{2,2} \mathcal{M}_{(2)}^2 + \boldsymbol{\gamma}_4 \mathcal{M}_{(4)} - \frac{1}{48} \mathcal{M}_{(2)}^3 \nonumber \\
&\quad + \frac{1}{8} \mathcal{M}_{(2)} \mathcal{M}_{(4)} - \frac16 \mathcal{M}_{(6)}\Bigr) 
\end{align} 
and the constants $\boldsymbol{\gamma}_i$ in \eqref{rel-O6-M} are fixed as
\begin{align}
\label{eq:gamma2-M}
\boldsymbol{\gamma}_2 &= -\frac{1}{12} \left(D^4 - 6 D^3 + 13 D^2 - 8 D - 4  \right), \\
\label{eq:gamma22-M}
\boldsymbol{\gamma}_{2,2} &=\frac{1}{48} \left(- 3 D^2 + D - 2  \right), \quad 
\boldsymbol{\gamma}_4 = \frac{1}{6} \left(2 D^2 - 5 D + 5 \right).
\end{align}

As expected the constants $\boldsymbol{\beta}_i$ from \eqref{eq:C6-AdS} match with \eqref{eq:beta-spin}, as it should be for the correct flat limit \eqref{flat-l-cas}. So the Casimir operators \eqref{eq:C2-AdS-D}-\eqref{eq:C6-AdS-D}  are  analogue for the $\mathbf{{S}_{p+1,q}}$ case of the Casimirs \eqref{eq:C2-iso}  and  \eqref{eq:C4-spin}-\eqref{eq:C6-spin} for the flat case\footnote{In this case generators of  $\mathfrak{iso}(p,q)$ have the form $P_n = \partial_n$ and $J_{m n}$ are taken from \eqref{ex-gen-sopq}. It also follows from the flat limit: $\mathcal{D}_n \overset{R \to \infty}{=} \partial_n$. }. The proof of \eqref{eq:C2-AdS-D} is cumbersome, but it can be done directly by hands. The Casimirs in the form \eqref{eq:C4-AdS-D}-\eqref{eq:C6-AdS-D} were found using $\mathfrak{Cadabra}$ \cite{cadKP} with help of Remark {\bf \ref{rep-for-cadab}}. 

\setcounter{equation}{0}
\section{Spinor formulation for the $\mathfrak{so}(2,5)$ case} 
Now we turn to the $\mathbf{AdS_6}$ case, which is the $\mathbf{S_{1+1,5}}$ space following our previous notation.
We will construct continuous spin representations of the $\mathfrak{so}(2,5)$ algebra  realized on the space of functions $\Psi(x^{\mu}, \rho^{A}_{\alpha})$, where $x^{\mu}$ are coordinates on $\mathbf{AdS_6}$,  with, $\mu = 0, \dots, 5$, and the $SU(2)$ Majorana-Weyl spinor $\rho^{A}_{\alpha}$ (see \cite{BFI-6D-tf} and refs. therein) is an auxiliary variable, which is related to the realization of the spin part $\mathcal{M}_{m n}$ of the generators $J_{m n}$ from  \eqref{s-f-1} and \eqref{ex-gen-sopq}. In what follows  we fix $m, n, k, l, ... = 0, 1, 2, \dots 5$ and $||\eta_{m n}|| = \mathrm{diag}(-,+,+,+,+,+)$. 

\subsection{Spinor notations}

A spinor realisation of the spin  part of the angular momentum $\mathcal{M}_{m n}$ from \eqref{ex-gen-sopq} can be represented in the following form 
\be \lb{Mreal}
\mathcal{M}_{m n} = - \rho^A_{\alpha} \, (\tilde{\sigma}_{m n})^{\alpha}_{\;\; \beta} \, \frac{\partial}{\partial \rho^{A}_{\beta}} \,
\ee
with the constraint 
\be \lb{spin-com}
\left[\frac{\partial}{\partial \rho^{A}_{\alpha}}, \rho^{B}_{\beta}\right] = \delta^{A}_{B} \delta_{\beta}^{\alpha} \, ,
\ee
where $A,B = 1,2 $ are $\mathfrak{su}(2)$ indices and $\alpha, \beta = 1,\dots,4$ are $\mathfrak{su}^*(4)$ indices.  In \eqref{Mreal} $(\tilde{\sigma}_{m n})^{\alpha}_{\;\; \beta}$ are defined as
\be \lb{sigmn}
(\tilde{\sigma}_{m n})^{\alpha}_{\;\; \beta} = \frac14 ( \tilde{\sigma}_m \sigma_n - \tilde{\sigma}_n \sigma_m)^{\alpha}_{\;\; \beta} \,, 
\ee
where for the $\sigma$-matrices  we used
$(\tilde{\sigma}_n)^{\alpha \beta} = - (\tilde{\sigma}_n)^{\beta \alpha}$ and $(\sigma_n)_{\alpha \beta} = - (\sigma_n)_{\beta \alpha}$ . 

\begin{remark}
 Futher, we will use various formulae for the $\sigma$-matrices. We use exactly the conventions proposed in \cite{BFI-6D-tf}, Appendix \textbf{A}. However, one needs to be careful, since in this  work we have the "mostly-plus" signature of the metric. 
\end{remark}

We introduce the following quantities using a combination of the spinors $\rho^{A}_{\alpha}$ and the differential operators $\frac{\partial}{\partial \rho^{A}_{\alpha}}$
\be \lb{N-and-bN}
N := \rho^{A}_{\alpha} \frac{\partial}{\partial \rho^{A}_{\alpha}}\,, \qquad \boldsymbol{N} := \rho^{A}_{\alpha} \frac{\partial}{\partial \rho^{B}_{\alpha}} \, \rho^{B}_{\beta} \frac{\partial}{\partial \rho^{A}_{\beta}} \,.
\ee
It is easy to see that the quantities $N$ and $\boldsymbol{N}$  commute $[N,\boldsymbol{N}] = 0$ as well as each of them commutes with $\mathcal{M}_{mn}$. In this sense we will call $N$ and $\boldsymbol{N}$ to be invariant.
A direct calculation of $\mathcal{M}_{(2)}$, defined as $\mathcal{M}_{(2)}=\mathcal{M}_{mn}\mathcal{M}^{nm}$ with \eqref{Mreal} leads to
\be \lb{M-sq}
\mathcal{M}_{(2)} = -\frac12 N^2 + 4 N + 2 \boldsymbol{N} \,.
\ee

By owning \eqref{N-and-bN} a bilinear symmetrized product of  $\mathfrak{so}(1,5)$ generators $\mathcal{M}_{mn}$ reads\footnote{Everywhere we use a standard normalization of symmetrization, i.e. $A_{(m} B_{n)} := \frac12 \left(A_m B_n + A_n B_m\right)$.}

\be \lb{sim-M-M}
\mathcal{M}_{(m}^{\;\;\;\;\;l} \mathcal{M}_{l n)} = \frac12 (\boldsymbol{N} + N - \frac12 N^2) \eta_{m n} - A_{(m} B_{n)} \,,
\ee
where vectors $A_m$ and $B_m$ are defined as follows
\be \lb{AB-def-n}
A_{m} = \frac12 \epsilon_{A B} \rho^{B}_{\alpha} \rho^{A}_{\beta} (\tilde{\sigma}_m)^{\alpha \beta} \,, \quad B_{m} = \frac12 \epsilon^{A B} \frac{\partial}{\partial \rho^{B}_{\alpha}} \frac{\partial}{\partial \rho^{A}_{\beta}} (\sigma_m)_{\alpha \beta} \, 
\ee
and $\epsilon_{A B}$ is a  standard $\mathfrak{su}(2)$ antisymmetric tensor with inverse $\epsilon^{A B}$.

Note that various commutators of the vectors $A_m$, $B_{m}$, the operators $\mathcal{M}_{mn}$ in the representation \eqref{Mreal} and covariant derivative $\mathcal{D}_{n}$ as well as their contractions have a number of useful relations, which we collect in App. \textbf{\ref{appSp}}.

\subsection{Casimir operators of $\mathfrak{so}(2,5)$ in spinor formulation}
In this section 
we aim to
 find the Casimir operators \eqref{eq:C2-AdS-D}-\eqref{eq:C6-AdS-D} in terms of the representation \eqref{Mreal} for $\mathcal{M}_{m n}$. First, we will derive new formulae for the invariant operators $(\mathcal{D}_{\!_{\bf{(1)}}})^2$, and $(\mathcal{D}_{\!_{\bf{(2)}}})^2$, which we introduced in \eqref{LBO-1}-\eqref{LBO-2}. For $(\mathcal{D}_{\!_{\bf{(1)}}})^2$  we obtain
\be \lb{D1inSp}
(\mathcal{D}_{_{\bf{(1)}}})^2 = - \frac12 (\boldsymbol{N} + N - \frac12 N^2) (\mathcal{D})^2 + (A^m \mathcal{D}_m) (B^n \mathcal{D}_{n}) - \frac{1}{2 R^2} A^m B^n \mathcal{M}_{n m} - \frac{1}{R^2} \mathcal{M}_{(2)} \,.
\ee
One can prove \eqref{D1inSp} performing the following steps. In \eqref{LBO-1} we split the contraction of $\mathcal{M}_{mn}$ to a symmetric and antisymmetric parts, i.e.  we have
\bea \lb{appB-1}
(\mathcal{D}_{_{\bf{(1)}}})^2 &=& -(\mathcal{M}^{(m k} \mathcal{M}_{k}^{\;\; n)} + \mathcal{M}^{[m k} \mathcal{M}_{k}^{\;\; n]}) (\mathcal{D}_{m} \mathcal{D}_{n} + w_{m n l} \mathcal{D}^l) \,\\
&=& - \mathcal{M}^{(m k} \mathcal{M}_{k}^{\;\; n)} (\mathcal{D}_{m} \mathcal{D}_{n} + w_{m n l} \mathcal{D}^l) - \frac{1}{R^2} \mathcal{M}_{(2)} \,, 
\eea
where we also used the relation for the Riemann and torsion tensors \eqref{tR-tT-ads} and the property of the covariant derivative \eqref{s-f-6D}. Now let us consider the first term in \eqref{appB-1} separately
\bea
\nonumber
- \mathcal{M}^{(m k} \mathcal{M}_{k}^{\;\; n)} (\mathcal{D}_{m} \mathcal{D}_{n} + w_{m n l} \mathcal{D}^l) &=& - \frac12 (\boldsymbol{N} + N - \frac12 N^2) (\mathcal{D})^2 + A^{m} B^n (\mathcal{D}_{(m} \mathcal{D}_{n)} + w_{(m n) l} \mathcal{D}^l) \\[7pt] 
 \lb{appB-2}
&=& - \frac12 (\boldsymbol{N} + N - \frac12 N^2) (\mathcal{D})^2 + (A^m \mathcal{D}_m) (B^n \mathcal{D}_n) \\[7pt] \nonumber
&-& \frac{1}{2R^2} A^m B^n \mathcal{M}_{n m},
\eea
where we used \eqref{sim-M-M}, \eqref{s-f-6D}, the second commutator from \eqref{AB-D-com} 
and again \eqref{tR-tT-ads}. Plugging \eqref{appB-2}  into \eqref{appB-1} we get the relation \eqref{D1inSp} for $(\mathcal{D}_{_{\bf{(1)}}})^2$.

As for $(\mathcal{D}_{_{\bf{(2)}}})^2$, doing some algebra we can come to
\bea \nonumber
(\mathcal{D}_{_{\bf{(2)}}})^2 &=& \mathcal{M}^{m}_{\;\; k} \mathcal{M}^{k l} \mathcal{M}_{l r} \mathcal{M}^{r n} (\mathcal{D}_{m} \mathcal{D}_n + w_{m n l} \mathcal{D}^l) + 20 (\mathcal{D}_{_{\bf{(1)}}})^2 + 2 \mathcal{M}_{(2)} (\mathcal{D})^2 \\[7pt]
\lb{D2inSp}
&+& \frac{1}{R^2}\left(20 \mathcal{M}_{(2)} - 2 \mathcal{M}_{(4)}\right)\,,
\eea
where we  used only the commutation relations for $\mathcal{M}_{m n}$.  Next, performing the transformations as above for $(\mathcal{D}_{_{\bf{(1)}}})^2$  
with the first term of \eqref{D2inSp} and taking into account the formulas \eqref{M4-sym} and \eqref{M5i}, we  are brought to
\bea \nonumber
(\mathcal{D}_{_{\bf{(2)}}})^2 &=& \mathcal{A}(\boldsymbol{N}, N) (\mathcal{D})^2 - \mathcal{B}(\boldsymbol{N},N) \left((A^m \mathcal{D}_m)(B^n \mathcal{D}_n) - \frac{1}{2 R^2} A^m B^n \mathcal{M}_{n m}\right) \\[7pt]
&+& 20 (\mathcal{D}_{_{\bf{(1)}}})^2 + 2 \mathcal{M}_{(2)} (\mathcal{D})^2  + \frac{1}{R^2} \left(\frac{3}{2} \mathcal{M}_{(4)} + 10 \mathcal{M}_{(2)} - \frac{1}{4} \mathcal{M}_{(2)}^2\right) \,,
\eea
where $\mathcal{A}(\boldsymbol{N}, N)$ and $\mathcal{B}(\boldsymbol{N}, N)$ are defined in \eqref{defAmc} and \eqref{defBmc}, respectively.

\begin{remark} \lb{r-loltl}
Using results this subsection we see that the Casimir operators \eqref{eq:C2-AdS-D}-\eqref{eq:C6-AdS-D} can be rewritten in terms of the  spinor invariants $ \boldsymbol{N}, N$ and the following three operators: $(\mathcal{D})^2\,,$ $ (A^m \mathcal{D}_m)$ and  $(B^m \mathcal{D}_m)$.
\end{remark}

\subsection{Continuous spin constraints}
In this subsection we will find constraints governing continuous spin field dynamics in the $\mathbf{AdS_6}$  space. Here, by constraints, we mean a system of differential relations, defined on the functions $\Psi(x^{\mu},\rho^{A}_{\alpha})$, which depend on the coordinate $x^{\mu}$ on $\mathbf{AdS_6}$ and the auxiliary spinor $\rho^A_{\alpha}$. That is, by the system of constraints for continuous spin in $\mathbf{AdS_6}$, we mean a set of such differential operators $\mathfrak{L}_i$ that consistently describe the dynamics of the continuous spin field propagating in $\mathbf{AdS_6}$. We will follow exactly the  paper \cite{BFIK-4D-ads}, except that in our work one of the constraints will be postulated. As we will discuss it below such assumption is quite natural.

As in the previous subsection, we mean everywhere the realization of $\mathcal{M}_{m n}$ in the form \eqref{Mreal}.

\paragraph{Constraints}
First, let us introduce the following operators\footnote{\label{footnote9}In work \cite{D6-LC} a set of the following operators was introduced: $\mathrm{l}_0 := \partial^{m}\partial_{m}$, $\tilde{\mathrm{l}} := A^{m}\partial_{m}-\boldsymbol{\mu}$, $\mathrm{l} := B^{m}\partial_{m}-\boldsymbol{\mu}$, $U$, where $A_m$, $B_m$ match with \eqref{AB-def-n}, $\boldsymbol{\mu}$ is a real parameter, $\partial_m := \partial/\partial x^m$, $x^m$ are coordinates on $\mathbb{R}^{1,5}$ and the operator $U$ is given by $\eqref{oU}$. The operators $\mathrm{l}_0, \,\tilde{\mathrm{l}}, \, \mathrm{l}, \, U$ form a closed algebra and give rise to constraints, which describe dynamics of continuous spin field in a six-dimensional flat space. To generalize the constraints to the $\mathbf{AdS_6}$ case we have to deform the operators consistently so that the deformed  operators  will form a closed algebra again. It turns out that for such deformation it is not enough to replace partial derivatives with the covariant ones; it is necessary to include a purely spinor contribution for each operator.  A detailed description of this procedure is given in work \cite{BFIK-4D-ads}.}  
\be \label{tildell}
\tilde{l} := (A^m \mathcal{D}_m) \,, \qquad l := (B^m \mathcal{D}_m) \,, 
\ee
where the vectors $A^{m}$, $B^{m}$ are defined in \eqref{AB-def-n} and $\mathcal{D}_m$ is given in \eqref{s-f-5} for $\mathbf{AdS_6}$.

For the above operators \eqref{tildell}  we have\footnote{A detailed calculation is given in App. \textbf{\ref{app:AppCR}}.}
\be \lb{inlpl}
[\tilde{l}, l] = K l_0 \,, 
\ee
where 
\be \lb{l0}
l_0 := (\mathcal{D})^2 + \frac{1}{R^2} \left (\frac12 \boldsymbol{N} - 2 N - \frac12 N^2  \right) \,
\ee
and the parameter $K=N+4$ was defined in \eqref{A-B-com}.

To proceed with finding the continuous spin constrains for the $\mathbf{AdS_6}$ case we perform a consistent deformation of the operators $l_0\,, \tilde{l} \,, l$ by the rules 

\bea \lb{newc}
l_0 \quad & \to & \quad L_0 = l_0 + \Delta_0\bigl(N, \boldsymbol{N}\bigr) \,, \\[8pt]
\lb{newcc}
\tilde{l} \quad & \to & \quad \tilde{L} = \tilde{l} + \Delta \bigl(N, \boldsymbol{N}\bigr) \,,  \\[8pt]
\lb{newccc}
l \quad & \to & \quad L = l + \Delta \bigl(N, \boldsymbol{N}\bigr) \,.
\eea 
with 
\be 
\Delta \bigl(N, \boldsymbol{N}\bigr) = - \boldsymbol{\mu} + \dots ,
\ee 
i.e. in \eqref{newc}-\eqref{newccc} we added a purely spinor terms $\Delta_0 \,, \Delta$, which, moreover,  depend on the parameter $R$. The later is related that under the flat limit $R \to \infty$,  we come to the constraints for continuous spin field in the 6d Minkowski spacetime studied in the work \cite{D6-LC} (see footnote {\bf \ref{footnote9}}).

To summarize we impose two requirements for such a deformation:
\begin{itemize}
     \item The operators $L_0 \,,  \tilde{L} \,, L$ form a closed algebra.  
    \item In the flat limit, i.e. when the parameter $R \to \infty$, we get the result of \cite{D6-LC} 
\end{itemize}

As we discussed above we change one of the steps in determining the correct constraint deformation. Namely, we assume that the operator $L_0$ has the following form
\be \label{L0}
L_0 = (\mathcal{D})^2 - \frac{1}{2 R^2} \mathcal{M}_{m n} \mathcal{M}^{m n} + a \,, 
\ee
where $a$ is some constant, which we will find below. 

The  requirement \eqref{L0} regarding the form of $L_0$ is related to certain results of the work \cite{BFIP},  the quadratic Casimir operator $\mathcal{C}_{2}$ in the form \eqref{eq:C2-AdS-D} and a  property of $L_0$ that we will present later.
Owning \eqref{newc} and \eqref{L0}, the quantity $\Delta_0$ reads
\be \lb{Del0}
\Delta_0 = a + \frac{1}{R^2} \left (4 N + \frac14 N^2 + \frac12 \boldsymbol{N} \right ) \,. 
\ee
Now let us consider the  commutator of $\tilde{L}$ and $L$
\be \lb{LpL}
[\tilde{L}, L] = K L_0 + (\check{\Delta}-\Delta) \tilde{L} - (\hat{\Delta}-\Delta)L - K \Delta_0 - (\check{\Delta} - \hat{\Delta}) \Delta \,,
\ee
where $\check{\Delta}$ and $\hat{\Delta}$ are defined by
\be \lb{Dchhat}
\check{\Delta} := \Delta\bigl(N-2, \boldsymbol{N}-2(N-1) \bigr) \,,\qquad \hat{\Delta} := \Delta\bigl(N+2, \boldsymbol{N}+2(N+1) \bigr) \,. 
\ee
These relations arise as a result of passing $A_m$ and $B_{m}$ through arbitrary polynomial functions of $N$ and $\boldsymbol{N}$, see \eqref{AmN-BmN}-\eqref{AmN-BmN-1}. It is worth to be noted that in \eqref{LpL} the last two terms appear because of completing the operators $l_0\,, \tilde{l}\,, l$ to the full ones $L_0\,, \tilde{L}\,, L$, i.e. according to the rules \eqref{newc}-\eqref{newccc}.

 Thus, the following equation can be read off from the requirement of closure of the algebra
\be \lb{eqonD}
K \Delta_0 + (\check{\Delta} - \hat{\Delta}) \Delta = 0.
\ee
Taking the ans\"atze for $\Delta$ in the form 
\be \lb{anonD}
\Delta = - \boldsymbol{\mu} + e + b\, N + c \, N^2 + d \, \boldsymbol{N}, \,
\ee
with   some parameters $b$,$c$,$d$,$e$
and plugging \eqref{anonD} back into \eqref{eqonD}, we find solutions for  $a,b,c,d,e$, namely
\be \lb{solonpar}
a = \mp \frac{2(e-\boldsymbol{\mu})}{R} \,,\quad b = \mp \frac{2}{R} \,,\quad c = \mp \frac{1}{8 R} \,,\quad d = \mp \frac{1}{4 R} \,,
\ee
where $e$ is arbitrary and we
recall, that the parameter $a$ appears in the definition of $\Delta_{0}$ \eqref{Del0}.
\paragraph{Full system of constraints} --
Without lost of generality,  below,  we will  focus on the  first solution from \eqref{solonpar}, thus the operators take the following form
\bea
\lb{oL0}
L_0 &=& (\mathcal{D})^2 - \frac{1}{2 R^2} \mathcal{M}_{m n} \mathcal{M}^{m n} - \frac{2(e- \boldsymbol{\mu})}{R} \,, \\[8pt]
\lb{oLt}
\tilde{L} &=& (A^m \mathcal{D}_m) - \boldsymbol{\mu} + e - \frac{2}{R} N - \frac{1}{8 R} N^2 - \frac{1}{4R} \boldsymbol{N} \,, \\[8pt] 
\lb{oL}
L &=& (B^m \mathcal{D}_m) - \boldsymbol{\mu} + e - \frac{2}{R} N - \frac{1}{8 R} N^2 - \frac{1}{4 R} \boldsymbol{N} \,, \\[8pt] 
\lb{oU}
U &=& \frac12\Bigl(\boldsymbol{N} - \frac{1}{2} N^2\Bigr) -  s(s+1) \,, 
\eea
where $s$ is a non-negative (half-) integer parameter.
The last constraint $U$ is the $\mathfrak{so}(3)$ Casimir operator and was found in the description of the continuous spin field in flat space in \cite{BFI-6D-tf}, however, this condition does not need to be deformed for the  $\mathbf{AdS}$ case. 

\paragraph{A simplified version of the system of constraints} -- Next we make some reasonable simplifications of \eqref{oL0}–\eqref{oL} to present them in a  compact form similar to that found in \cite{BFIK-4D-ads} for the $\mathbf{AdS_4}$  case. To achieve the required form, first, we use the constraint \eqref{oU} in \eqref{oLt}-\eqref{oL} and, then, we fix parameter $e$ as $e = -4/R$. Finally, one gets
\bea
\lb{nL0}
L_0 &=& (\mathcal{D})^2 - \frac{1}{2 R^2} \mathcal{M}_{m n} \mathcal{M}^{m n} + \frac{2(4+ \boldsymbol{\mu} R)}{R^2} \,, \\[8pt]
\lb{nLt}
\tilde{L} &=& (A^m \mathcal{D}_m) - \boldsymbol{\mu} - \frac{s(s+1)}{2R}  - \frac{K^2}{4R} \,, \\[8pt]
\lb{nL}
L &=& (B^m \mathcal{D}_m) - \boldsymbol{\mu} - \frac{s(s+1)}{2R}  - \frac{K^2}{4R}  \,.
\eea

\paragraph{Algebra of constraints} -- The commutators of the constructed operators $L_{0}$, $\tilde{L}$, $L$ read
\be \lb{g-a-wlpl}
[L_0, \tilde{L}] = [L_0, L] = [L_0, U] = 0 \,, \qquad [U, \tilde{L}] = [U, L] = 0 \,,
\ee
\be \lb{ga-lpl}
[\tilde{L}, L] = K L_0 + R^{-1} \left(K - 1\right) \tilde{L} + R^{-1} \left( K + 1 \right) L \,.
\ee
The  commutator \footnote{Compare this with the corresponding expression from the work \cite{BFIK-4D-ads}.} given by \eqref{ga-lpl} immediately follows from \eqref{LpL} with the parameters \eqref{solonpar} and fixing $e=-4/R$. The right chain of equalities in \eqref{g-a-wlpl} is a sequence of eqs. \eqref{AmN-BmN}-\eqref{AmN-BmN-1}. Finally, the left chain of equalities in \eqref{g-a-wlpl} follows from the structure of the operator $L_0$. The later is another argument in favor of using the assumption \eqref{L0} . 
We will formulate it as remark.
\begin{remark}
Let us consider the  operator $\boldsymbol{X} = \alpha \, (\Sigma^n \mathcal{D}_n ) + F \left (N, \boldsymbol{N} \right)$, where $\alpha$ is an arbitrary constant, $\Sigma^n$ is an arbitrary vector, which does not depend on the coordinates $x^{\mu}$ of the space and is transformed only by generators $\mathcal{M}_{m n}$ according to the rule of type \eqref{M-A-B-com}, $F$ is an arbitrary function of $N$ and $\boldsymbol{N}$. Then the equality holds
\be \lb{prop-1}
[L_0, \boldsymbol{X}] = 0 \,. 
\ee
\end{remark}

\begin{prop} The operator constraints $L_0\,, \tilde{L}\,, L\,, U$ define a representation of the Lie algebra $\mathfrak{so}(p+1,q)$.
\\
\textbf{Proof:} The operator constraints $L_0\,, \tilde{L}\,, L\,, U$ commute with the generators\footnote{Here we mean the explicit realization of the generators for $\mathfrak{so}(p+1,q)$ found in subsection \ref{susec-gen} (see especially Remarks {\bf \ref{rem-1}-\ref{rem-2}}), for $p =1, q=5$ and the spin part of the angular momentum $\mathcal{M}_{m n}$ defined in \eqref{Mreal}.} $P_n$ and $J_{m n}$ of the $\mathfrak{so}(2,5)$ algebra. 
\end{prop}

In addition to the fact that the operator constraints found determine the representation of the symmetry algebra, they also resolve all Casimir operators. In the sense that functions satisfying these constraints are eigenfunctions of the Casimir operators (a similar statement is given  for the $\mathbf{AdS_{4}}$ space in \cite{BFIP}). Below we consider this fact in more detail.

\paragraph{Resolution of the Casimir operators} --
Let the function $\Psi(x^{\mu},\rho^{A}_{\alpha})$ satisfy the system of differential equations
\be \lb{constr-one-l}
L_0 \Psi = 0\,, \quad \tilde{L} \Psi = 0 \,, \quad L \Psi = 0\,, \quad U \Psi = 0 \,,
\ee
where under $L_0\,, \tilde{L}\,, L$ we mean  \eqref{nL0}-\eqref{nL} and  $U$ is given by \eqref{oU}.

Then taking into account Remark \ref{r-loltl} we can use these constraints  for the Casimir operators of the $\mathbf{AdS_6}$ case given by \eqref{eq:C2-AdS-D}-\eqref{eq:C6-AdS-D} . It turns out that the functions $\Psi(x^{\mu},\rho^{A}_{\alpha})$ satisfying  the equations \eqref{constr-one-l} are eigenfunctions for the Casimir operators with the following eigenvalues 
\bea
\lb{eq:evC2}
\mathcal{C}_2 &\simeq& 2(4+\boldsymbol{\mu} R)\,,\\[7pt] 
\lb{eq:evC4}
\mathcal{C}_4 &\simeq&  \left(\boldsymbol{\mu} R - 3 S + 4\right) \left(\boldsymbol{\mu} R+S+1\right)\,, \\[7pt]
\lb{eq:evC6}
\mathcal{C}_6 &\simeq& - 2 S \left(\boldsymbol{\mu}^2 R^2 +  (2S+1) \boldsymbol{\mu} R+ S(S+1) \right)\,,
\eea
where $S = s(s+1)/2$ and the sign $\simeq$ means that the Casimirs are calculated under the condition \eqref{constr-one-l}. In the flat limit \eqref{flat-l-cas} eqs. \eqref{eq:evC2}-\eqref{eq:evC6} has form 
\be \lb{flacas}
C_2 \simeq 0 \,, \quad C_4 \simeq \boldsymbol{\mu}^2 \,, \quad C_6 \simeq - \boldsymbol{\mu}^2 s (s + 1) \,,
\ee
which, up to the sign of the second Casimir\footnote{For a complete match, the following redefinitions of the $\mathfrak{so}(2,5)$ generators and flat metric must be made: $P_n \to i P_n, \;\; J_{m n} \to i J_{m n}, \;\; \eta_{m n} \to - \eta_{m n}$.}, coincides with the results of works \cite{BFI-6D-tf, D6-LC}.

\begin{remark} \lb{M-Cl}
    A straightforward analysis based on eqs.~\eqref{eq:evC2}-\eqref{eq:evC4} (see also \cite{BFIP}) shows that our constructed representation with $s=0$ corresponds to the following cases in Metsaev's classification \cite{l-3-5}:
\begin{itemize}
    \item Type $\mathbf{ii}$ and $\mathbf{iii}$ for $\boldsymbol{\mu}>0$
\end{itemize}
\end{remark}

\begin{remark} The constraints $L_0\,, \tilde{L}\,, L\,, U$ form the algebra \eqref{g-a-wlpl}-\eqref{ga-lpl}, which is similar to the algebra found in \cite{BFIK-4D-ads}. It is interesting, that if we can try to find the constraints like $L_0\,, \tilde{L}\,, L\,, U$  directly from the Casimir operators (as the conditions for its resolution) we will see that there would be one free parameter, which can only be fixed by the condition of the closure of the algebra of these constraints
\end{remark}

\section*{Discussion} 
\addcontentsline{toc}{section}{Discussion}
Now we summarize  our main foundings. In this work we have  investigated
continuous spin representation of the  isometry group of six-dimensional anti-de Sitter space $\mathbf{AdS_6}$, corresponding to the Lie algebra $\mathfrak{so}(2,5)$.

First, we chose to realize the anti-de Sitter space $\mathbf{AdS_6}$  as a hypersurface $\mathbf{S_{1+1,5}}$ embedded into  $\mathbb{R}^{1+1,5}$.
Then by making a
stereographic projection of the hypersurface, the generators of $\mathfrak{so}(2,5)$ were presented in terms of geometric quantities.
We observed that the generators of $\mathfrak{so}(2,5)$ in the representation under consideration can be expressed through the Lie-Lorentz derivative.

For the spinor formulation of  $\mathfrak{so}(2,5)$ we have constructed operator constraints, that define the continuous spin field in the $\mathbf{AdS_6}$ space. The constraints depend on the following three parameters: the $\mathbf{AdS}$ radius $R$, a real parameter $\boldsymbol{\mu}$, and a positive (half-)integer $s$. As the radius approaches infinity, one comes to the flat limit and the derived constraints reproduce the known results for the $6d$ Minkowski space. We have shown that these constraints realize  the $\mathfrak{so}(2,5)$ representation, characterized by $\boldsymbol{\mu}$ and $s$, and fix the eigenvalues of all Casimir operators. For the simplest case $s=0$, a comparison with Metsaev's classification was also carried out in Remark {\bf \ref{M-Cl}}.

Although in this paper our primary emphasis has been on representations of continuous spin fields in the $\mathbf{AdS}$ space of a specific dimension, we have obtained a number of results for more general curved spaces. Particularly, in the  case of  the pseudosphere $\mathbf{S_{p+1,q}}$  we have demonstrated that the corresponding Casimir operators can be expressed through the covariant derivative and the spin part of the angular momentum.

There are several outlooks for this work.
For the future it would be interesting to prove unitarity for the constructed representation. Another extension is the construction of a BRST Lagrangian for $\mathbf{AdS_6}$ continuous spin field, which requires further investigation. 

It might also be of interest to
 generalize this approach to the case of continuous spin field in the $\mathbf{AdS_D}$ space. This seems possible since, judging by the cases of $\mathbf{AdS_{4}}$ and $\mathbf{AdS_{6}}$ cases, the procedure for constructing the operator constraints $L_{0}, \, \tilde{L}$ and $L$ does not become more complicated with increasing dimension. However, finding  higher-order constraints and the direct check that the resulting constraints fix all Casimir operators requires some modification.

\section*{Acknowledgments}

 It is a pleasure to thank  I.L. Buchbinder, A.P. Isaev and S.A. Fedoruk for useful discussions at the early stage of this project. We especially thank E. Skvortsov for comments and feedback on the manuscript. M.A.P. is grateful to S.O. Krivonos for showing useful Wolfram Mathematica techniques that helped in this project. 

\appendix

\section{Some general geometric notation}
\label{appGeom}
\setcounter{equation}{0}\renewcommand{\theequation}{A.\arabic{equation}}

In this Appendix we recall useful general geometric  quantities (see for e.g. books  \cite{IR, BK})
\paragraph{Frame fields and anholonomy coefficients} --
Let $g_{\mu \nu}(x)$ be a metric tensor of a $D$-dimensional (pseudo) Riemannian manifold, so $g_{\mu \lambda}(x) g^{\lambda \nu}(x) = \delta_{\mu}^{\nu}$, where $g^{\mu \nu}(x)$ is an inverse metric. The frame fields $e^{\mu}_{n}$ can be defined by relations\footnote{We associate Latin indices with a flat space, and Greek indices with a Riemannian manifold.}
\be \lb{FrFi}
e_{\mu}^{n} e_{\nu}^{m} \eta_{n m} = g_{\mu \nu}\,, \quad e^{\mu}_{n} e^{\nu}_{m} g_{\mu \nu} = \eta_{n m} \,,
\ee
where $\mu, \nu = 0, 1, \dots D-1$; $m, n = 0,1, \dots , D-1$ and 
\be \lb{flat-metric}
||\eta_{n m}|| = \mathrm{diag}(\underbrace{-1, \dots, -1}_{p}, \underbrace{+1, \dots, +1}_{q}) \, .
\ee
Let us recall the notion of the anholonomy tensor 
\be \lb{AnhC}
[e_n, e_m] = \mathcal{E}_{n m}^{\;\;\;\;\;l} e_l \,,
\ee
where we define 
\be \lb{frf}
e_{n} := e_{n}^{\mu} \partial_{\mu} \,, \quad  \mathcal{E}_{n m}^{\;\;\;\;\;l} := (e_n e_{m}^{\mu} - e_m e_n^{\mu}) e_{\mu}^l \,, 
\ee
the quantity $\mathcal{E}_{n m}^{\;\;\;\;\;l}$ is the anholonomy tensor. 

\paragraph{Riemann and torsion tensors} --
Let us also introduce the spin connection
$w_{k n m} = - w_{k m n}$ and the  torsion tensor $\mathcal{T}_{n m}^{\;\;\;\;\;l}$ 
\be \lb{totT}
\mathcal{T}_{n m}^{\;\;\;\;\;l} = \mathcal{E}_{n m}^{\;\;\;\;\;l} + w_{n m}^{\;\;\;\;\;l}-w_{m n}^{\;\;\;\;\;l} \,. 
\ee
The Riemann tensor $\mathcal{R}_{n m}^{\;\;\;\;\;kl}$  is given by\footnote{We assume throughout that the spin-connection is antisymmetric in the last two indices, and the anholonomy tensor is antisymmetric in the first two indices.} 
\be \lb{RtCA}
\mathcal{R}_{n m}^{\;\;\;\;\;k l} = e_n w_m^{\;\;kl}-e_m w_n^{\;\;kl} - \mathcal{E}_{nm}^{\;\;\;\;\;r} w_r^{\;\;kl}+ w_n^{\;\;k r} w_{m r}^{\;\;\;\;l} -  w_m^{\;\;k r} w_{n r}^{\;\;\;\;l}.
\ee
In the torsion free case, i.e. $\mathcal{T}_{n m}^{\;\;\;\;\;l}=0$, using \eqref{totT} we get the relation 
\be \lb{sc-ac}
w_{k n m } = \frac{1}{2} (\mathcal{E}_{ n m k} + \mathcal{E}_{k m n} - \mathcal{E}_{ k n m} ) \,.
\ee

\section{Useful spinor relations}
\label{appSp}
\setcounter{equation}{0}\renewcommand{\theequation}{B.\arabic{equation}}
In this appendix we collect useful spinor relations.
\\

The objects $A_m$ and $B_m$ given by \eqref{AB-def-n} are vectors relative to the $\mathfrak{so}(1,5)$ generators $\mathcal{M}_{m n}$

\be \lb{M-A-B-com}
[\mathcal{M}_{m n}, A_l] = \eta_{n l} A_{m} - \eta_{m l} A_{n} \,, \qquad [\mathcal{M}_{m n}, B_l] = \eta_{n l} B_{m} - \eta_{m l} B_{n} \,. 
\ee 

Various commutators between the quantities $A_m, B_m$ and $N, \boldsymbol{N}$ can be simplified using the following relations

\be \lb{A-B-com}
[A_m, B_n] = K \eta_{m n} + 2 \mathcal{M}_{m n} \,, \qquad K := (N+4) \,, 
\ee
\bea \lb{AmN-BmN}
A_{m} \, F\bigl(N, \boldsymbol{N}\bigr) &=& F\bigl(N-2,\boldsymbol{N}-2(N-1)\bigr) \, A_{m}, \\[7pt]
\lb{AmN-BmN-1}
B_{m} \, F\bigl(N, \boldsymbol{N}\bigr) &=& F\bigl(N+2,\boldsymbol{N}+2(N+1)\bigr) \, B_{m} \,,
\eea
where $F$ is an arbitrary polynomial function of $N$ and $\boldsymbol{N}$.

We also actively use the following commutation relations
\be\lb{AB-D-com}
[\mathcal{D}_n, A_{m}] = - w_{n m l} A^l \,, \qquad [\mathcal{D}_n, B_{m}] = - w_{n m l} B^l\,,
\ee
where we take into account \eqref{s-f-5} and \eqref{M-A-B-com}. 

Various contractions of $A_m, B_n$ and $\mathcal{M}_{m n}$ are given below
\be \lb{eq:ABandBA}
A_m B^m = (\boldsymbol{N} - N -N^2) \,, \quad B_m A^m = (\boldsymbol{N} - 7 N - N^2 - 24 )\,,
\ee
\be \lb{eq:MBandMA}
\mathcal{M}_{m n} B^n = -\frac12 N B_{m} \,, \quad \mathcal{M}_{m n} A^{n} = \left(\frac12 N + 4 \right) A_{m} \,,
\ee
\be \lb{BAM-c}
B^n A^m \mathcal{M}_{m n} =  - \frac12 N (\boldsymbol{N}  - N^2 - 7 N - 24) \,, \quad A^m B^n \mathcal{M}_{n m} = \left (\frac12 N + 4\right) (\boldsymbol{N} - N - N^2) \,. 
\ee

A higher-order analog of the formula \eqref{sim-M-M} has the form 
\be \lb{M4-sym}
\mathcal{M}_{(l k} \mathcal{M}^{k n} \mathcal{M}_{n r} \mathcal{M}^{r}_{\;\;m)} = \mathcal{A}(\boldsymbol{N}, N) \eta_{l m} - \mathcal{B}(\boldsymbol{N},N) A_{(l} B_{m)} \,, 
\ee
where 
\be \lb{defAmc}
\mathcal{A}(\boldsymbol{N},N) := \left (\frac14 \boldsymbol{N}^2 + \frac72\boldsymbol{N} - \frac14 \boldsymbol{N} N^2 + \frac14 \boldsymbol{N} N + \frac{1}{16} N^4 - 2 N^2 + \frac12 N\right ) 
\ee
and
\be \lb{defBmc}
\mathcal{B}(\boldsymbol{N},N) :=  \left ( \frac{1}{2} \boldsymbol{N} + 2 N + 21 \right).
\ee
Eq.\eqref{M4-sym} was found using the following relation 
\bea
\mathcal{M}_{l k} \mathcal{M}^{k n} \mathcal{M}_{n r} \mathcal{M}^{r}_{\;\;m} &=& \mathcal{M}_{(l k} \mathcal{M}^{k n)} \mathcal{M}_{(n r} \mathcal{M}^{r}_{\;\;m)} + 4 \mathcal{M}_{l}^{\;\;n} \mathcal{M}_{(n r} \mathcal{M}^{r}_{\;\;m)} \\[7pt]
&+& 4 \mathcal{M}_{l k} \mathcal{M}^{k}_{\;\; m}. \nonumber
\eea
Thus for $\mathcal{M}_{(4)}$ we have 
\be \lb{eq::M4}
\mathcal{M}_{(4)} = \boldsymbol{N}^2 - \boldsymbol{N} N^2 + \frac{3}{8} N^4 + 2 N^3 + 11 N^2 + 24 N .
\ee
Below we also need to calculate $\mathcal{M}_{(6)}$. It can be done using the relation
\be
\mathcal{M}_{(6)} = \mathcal{M}_{(l}^{\;\;\;\;\;c} \mathcal{M}_{c a)} \mathcal{M}^{(l}_{\;\;b}\mathcal{M}^{b k} \mathcal{M}_{k d} \mathcal{M}^{d a)} + 2 \mathcal{M}_{(5)} \,, 
\ee
where $\mathcal{M}_{(5)}$ is defined similarly to the 
relations \eqref{eq:Jk-def} and we have the following representation 
\be \lb{M5i}
\mathcal{M}_{(5)} = 7 \mathcal{M}_{(4)} - 20 \mathcal{M}_{(2)} - \frac12 \left( \mathcal{M}_{(2)}\right)^2.
\ee
So for $\mathcal{M}_{(6)}$ we have
\bea
\nonumber
\mathcal{M}_{(6)} &=& \frac{\boldsymbol{N}^3}{2}-\frac{3 \boldsymbol{N}^2 N^2}{4}+13 \boldsymbol{N}^2+\frac{3 \boldsymbol{N} N^4}{8}-\frac{45 \boldsymbol{N} N^2}{2}-76 \boldsymbol{N} N-248 \boldsymbol{N} \\[7pt] \lb{eq::M6}
&-&\frac{N^6}{32} + \frac{3 N^5}{4} + \frac{71 N^4}{4}+114 N^3+407 N^2+344 N.
\eea
\paragraph{Relation between Casimirs} -- We can calculate $\mathfrak{C}_4$ and $\mathfrak{C}_6$ by formulae \eqref{eq:C4}-\eqref{eq:C6} using  operators $\mathcal{M}_{(2i)}$ from \eqref{M-sq}, \eqref{eq::M4}, \eqref{eq::M6} instead of $J_{(2i)}$. So, these $\mathfrak{so}(1,5)$ Casimir operators satisfy 
\be 
\mathfrak{C}_{4} \Bigl |_{\boldsymbol{N} = \frac12 N^2} = \mathfrak{C}_6 \Bigl |_{\boldsymbol{N} = \frac12 N^2} = 0.
\ee
Note, that constraints like $\boldsymbol{N} - \frac{1}{2} N^2 = \mathrm{const}$ were considered in works \cite{BFI-6D-tf}-\cite{D6-LC}.

\section{Commutator of $\tilde{l}$ and $l$}\label{app:AppCR}
\setcounter{equation}{0}\renewcommand{\theequation}{D.\arabic{equation}}

In this appendix  we discuss  relations, which are useful for finding operator constraints.

Consider the following commutator

\be \lb{lpl-com-ch}
[\tilde{l}, l] = 
(B^n A^m [\mathcal{D}_{m}, \mathcal{D}_n] + B^n[A^m,\mathcal{D}_n] \mathcal{D}_m + [A^m,B^n] \mathcal{D}_m \mathcal{D}_n + A^m [\mathcal{D}_m, B^n] \mathcal{D}_n)
\ee
Let us now find all commutators from \eqref{lpl-com-ch}  separately. The first commutator reads
\be \lb{A2}
B^n A^m [\mathcal{D}_m, \mathcal{D}_n] = B^n A^m \left(\mathcal{E}_{m n}^{\;\;\;\;\; l} \mathcal{D}_l + \frac12\mathcal{R}_{m n}^{\;\;\;\;\;k l} \mathcal{M}_{k l} \right)  = B^n A^m \mathcal{E}_{m n}^{\;\;\;\;\; l} \mathcal{D}_l - \frac{1}{R^2} B^n A^m \mathcal{M}_{m n} \,,
\ee
where we used eqs. \eqref{s-f-6D} and \eqref{tR-tT-ads}. 

The second one from \eqref{lpl-com-ch} can be represented as 
\be \lb{A3}
B^n[A^m,\mathcal{D}_n] \mathcal{D}_m = - B^n A^m w_{n m l} \mathcal{D}^{l}\,,
\ee
where we used \eqref{AB-D-com}.
The third term from \eqref{lpl-com-ch} can be brought to the following form
\be \lb{A4}
[A^m,B^n] \mathcal{D}_m \mathcal{D}_n = K \mathcal{D}_n \mathcal{D}^n + \mathcal{M}^{m n} \left(\mathcal{E}_{m n}^{\;\;\;\;\; l} \mathcal{D}_l + \frac12\mathcal{R}_{m n}^{\;\;\;\;\;k l} \mathcal{M}_{k l} \right) \,,
\ee
where we used \eqref{A-B-com} and \eqref{s-f-6D} again.

The last commutator comes to
\be \lb{A5}
A^m [\mathcal{D}_m, B^n] \mathcal{D}_n = A^m B^n w_{m n l} \mathcal{D}^l \,
\ee
where we used \eqref{AB-D-com} again. 
By now combining the first term from \eqref{A2} and \eqref{A3}, we obtain 
\be \lb{A6}
B^n A^m (\mathcal{E}_{m n}^{\;\;\;\;\; l} \mathcal{D}_l - w_{n m l} \mathcal{D}^{l}) = -B^n A^m w_{m n l} \mathcal{D}^l \,
\ee
where  eqs. \eqref{totT} and \eqref{tR-tT-ads} are taken into account. 
Combining \eqref{A6} with \eqref{A5} we get
\be \lb{A7}
(A^m B^n - B^n A^m) w_{m n l} \mathcal{D}^l = K w_{n}^{\;\; n l} \mathcal{D}_l + 2 \mathcal{M}^{m n} w_{m n l} \mathcal{D}^l\,
\ee
where we used \eqref{A-B-com}. 
Now we have to sum the second term of \eqref{A2}, all terms of \eqref{A4} and \eqref{A7}, so we get
\be \lb{A8}
[\tilde{l},l] = \left( K (\mathcal{D})^2 - \frac{1}{R^2} \left( B^n A^m \mathcal{M}_{m n} + \mathcal{M}_{m n} \mathcal{M}^{m n} \right ) \right)
\ee
where we used the definition of the operator $(\mathcal{D})^2$ given by \eqref{LBO-0} and relation
\be \lb{A9}
\mathcal{M}^{m n} \left (\mathcal{E}_{m n l} + w_{m n l} - w_{n m l}\right) \mathcal{D}^{l} = 0 \,
\ee
which is exactly the torsion-free constraint, i.e. eqs. \eqref{totT} and \eqref{tR-tT-ads}.

Finally, taking into account the relationns 
\eqref{M-sq} and \eqref{BAM-c} for the commutator  \eqref{A8} we are brought to \eqref{inlpl}-\eqref{l0}.

\end{document}